\documentclass[12pt]{elsart}

\usepackage{amsmath,amssymb,amsfonts}
\usepackage[mathcal]{euscript}
\usepackage{graphicx}
\usepackage{cite}

\newcommand{\eqlabel}[1]{\label{eq:#1}}

\newcommand{\seclabel}[1]{\label{sec:#1}}

\renewcommand{\eqref}[1]{(\ref{eq:#1})}

\newcommand{\secref}[1]{Section~\ref{sec:#1}}
\newcommand{\secreftwo}[2]{Sections~\ref{sec:#1} and~\ref{sec:#2}}

\newcommand{\mathnotation}[2]{\newcommand{#1}{\ensuremath{#2}}}

\renewcommand{\l}{\left}			
\renewcommand{\r}{\right}			
\mathnotation{\pd}{\partial}			
\newcommand{\holdconst}[1]{\Bigr|_{#1}}		
\newcommand{\sholdconst}[1]{|_{#1}}		
\mathnotation{\ee}{{\mathrm e}}			
\mathnotation{\ldef}{\mathrel{\raisebox{.069ex}{:}\!\!=}}
\mathnotation{\rdef}{\mathrel{=\!\!\raisebox{.069ex}{:}}}
\mathnotation{\levicivita}{\varepsilon}		

\mathnotation{\grad}{\nabla}			
\mathnotation{\lapl}{\nabla^2}			

\mathnotation{\covd}{\grad}			
\mathnotation{\Dt}{\mathcal{D}}			
\mathnotation{\Dtconv}{\Dt_{\mathrm{c}}}	
\mathnotation{\Dtdir}{\Dt_{\vel}}		
\mathnotation{\Dtcor}{\Dt_{J}}			
\mathnotation{\ptconnex}{\alpha}		
\mathnotation{\Connex}{\Gamma}			

\mathnotation{\dint}{\,{\mathrm{d}}}		
\mathnotation{\pheq}{&\phantom{=}}		
\renewcommand{\time}{t}				
\mathnotation{\lyapexp}{\lambda}		
\mathnotation{\lyapexpinf}{\lyapexp^\infty}	
\mathnotation{\x}{x}				
\mathnotation{\xv}{\x}				
\mathnotation{\velc}{v}				
\mathnotation{\vel}{\velc}			
\mathnotation{\Vel}{V}				
\mathnotation{\flow}{\varphi}			
\mathnotation{\Manif}{{\mathcal{U}}}		
\mathnotation{\sdim}{n}				
\mathnotation{\Tangent}{\mathrm{T}}		
\mathnotation{\Diff}{D}				
\mathnotation{\vvec}{X}				
\mathnotation{\wvec}{Y}				
\mathnotation{\dvdx}{G}				
\mathnotation{\G}{\dvdx}
\mathnotation{\A}{A}
\mathnotation{\lagrc}{a}			
\mathnotation{\lagrcv}{\lagrc}			
\mathnotation{\zv}{z}				
\mathnotation{\z}{z}				
\mathnotation{\metric}{g}			
\mathnotation{\anytens}{\mathcal{H}}		
\mathnotation{\anytenssym}{\anytens^{\mathrm{S}}}
\mathnotation{\anytensasym}{\anytens^{\mathrm{A}}}
\mathnotation{\arbtens}{\mathcal{T}}		

\mathnotation{\rostrain}{\gamma}		
\mathnotation{\Vortens}{\omega}			
\mathnotation{\rocstrain}{\kappa}		
\mathnotation{\Rcurv}{R}			
\mathnotation{\Scurv}{S}			
\mathnotation{\Bf}{B}				

\newcommand{\rostr}{rate-of-strain}
\newcommand{\crostr}{coordinate rate-of-strain}

\begin{document}

\journal{Journal of Physics A}
\volume{34}
\issue{29}
\pubyear{2001}

\begin{frontmatter}

\title{Covariant Time Derivatives for\\ Dynamical Systems}
\author{Jean-Luc Thiffeault}

\address{Department of Applied Physics and Applied Mathematics\\
	Columbia University, New York, NY 10027, USA}

\ead{jeanluc@mailaps.org}


\begin{abstract}

We present a unified derivation of covariant time derivatives, which transform
as tensors under a time-dependent coordinate change.  Such derivatives are
essential for formulating physical laws in a frame-independent manner.  Three
specific derivatives are described: convective, corotational, and directional.
The covariance is made explicit by working in arbitrary time-dependent
coordinates, instead of restricting to Eulerian (fixed) or Lagrangian
(material) coordinates.  The commutator of covariant time and space
derivatives is interpreted in terms of a \emph{time-curvature} that shares
many properties of the Riemann curvature tensor, and reflects nontrivial
time-dependence of the metric.

\end{abstract}

\end{frontmatter}

PACS numbers: 83.10.Bb, 05.45.-a, 47.50.+d

\vspace*{-1em}

\section{Introduction}

In physics, choosing an appropriate coordinate system can make the difference
between a tractable problem and one that defies analytical study.  In fluid
dynamics, two main types of coordinates are used, each representing a natural
setting in which to study fluid motion: the Eulerian coordinates, also known
as the laboratory frame, are time-independent and fixed in space; in contrast,
the Lagrangian (or material) coordinates are constructed to move with fluid
elements.  In between these extremes, other types of coordinates are used,
such as rotating coordinates in geophysical fluid dynamics.  Such coordinates
usually have a nontrivial spatial and temporal dependence.

The situation becomes more complicated when dealing with moving surfaces: here
the metric itself has intrinsic time dependence.  This time dependence
incorporates the strain imposed on a 2D surface flow as the surface deforms.
To properly formulate fluid equations on thin films and other surfaces, one
needs a \emph{covariant} description, that is, a description of the building
blocks of equations of motion---spatial and temporal derivatives---that obey
tensorial transformation laws.

There are other reasons than inherent deformation of the space to introduce a
time-dependent, nontrivial metric.  For instance, the advection-diffusion
equation can have an anisotropic, time-dependent diffusion tensor, perhaps
arising from some inhomogeneous turbulent process.  In that case, it is
advantageous to use the diffusion tensor as a metric, for then the
characteristic directions of stretching, given by the eigenvectors of the
metric tensor in Lagrangian coordinates, correspond to directions of
suppressed or enhanced diffusion associated with positive or negative Lyapunov
exponents, respectively~\cite{Thiffeault2001d,Thiffeault2001f}.

Local physical quantities can be viewed as tensors (scalars, vectors, or
higher-order tensors) evaluated along fluid trajectories.  For instance, we
may be interested in how the temperature (scalar) of a fluid element varies
along a trajectory, or how the magnetic field (vector) associated with a fluid
element evolves.  Characterising the evolution of these tensors in complicated
coordinates is again best done using some form of covariant time derivative,
also called an \emph{objective} time derivative.

The covariant spatial derivative is a familiar tool of differential
geometry~\cite{Misner,Schutz,Wald}.  The emphasis is usually on covariance
under coordinate transformations of the full space-time.  In fluid dynamics
and general dynamical systems, however, the time coordinate is not included in
the metric (though the metric components may depend on time), and the required
covariance is less restrictive: we seek covariance under time-dependent
transformations of the coordinates, but the new time is the same as the old
and does not depend on the coordinates.  Time derivatives lead to
non-tensorial terms because of time-dependent basis vectors---the same reason
that ordinary derivatives are not covariant.

There are many ways of choosing a covariant time derivative.  The most
familiar is the \emph{convective} derivative introduced by
Oldroyd~\cite{Oldroyd1950,Aris} in formulating rheological equations of state.
This derivative was then used by Scriven~\cite{Scriven1960} to develop a
theory of fluid motion on an interface.  The convective derivative of a tensor
is essentially its Lie derivative along the velocity vector.  In spite of its
economical elegance, the convective derivative has drawbacks.  Firstly, unlike
the usual covariant spatial derivative, it is not \emph{compatible} with the
metric tensor.  A compatible operator vanishes when acting on the metric.
Because the covariant derivative also has the Leibniz property, compatibility
allows the raising and lowering of indices ``through'' the operator.  This is
convenient for some applications~\cite{Thiffeault2001d}, and implies that the
equation of motion for a contravariant tensor has the same form as the
covariant one.  A second drawback of the convective derivative is that it
involves gradients of the velocity, and so is not \emph{directional}.  The
commutator of the convective derivative and the spatial derivative thus
involves \emph{second} derivatives of the velocity, requiring it to be at
least of class~$C^2$.

A second common type of derivative is the \emph{corotational} or
\emph{Jaumann} derivative (See Refs.~\cite{Jou} and~\cite[p.~342]{Bird1}, and
references therein), where the local vorticity of the flow is incorporated
into the derivative operator.  The corotational derivative is compatible with
the metric, but like the convective derivative it depends on gradients of the
velocity.

The third type of derivative we discuss is a new, time-dependent version of
the usual \emph{directional} derivative along a curve used to define parallel
transport~\cite{Misner,Schutz,Wald}.  The curve here is the actual trajectory
of a fluid particle, with tangent vector given by the Eulerian velocity field.
The directional derivative does not depend on gradients of the velocity field.
The concept of time-dependent parallel transport can be introduced using this
derivative, and is equivalent to a covariant description of advection without
stretching.  A directional derivative was introduced in the context of fluid
motion by Truesdell~\cite[p.~42]{Truesdell}, but it does not allow for
time-dependence in the coordinates or metric.  (Truesdell calls it the
\emph{material} derivative because of its connexion to fluid elements.)

In this paper, we present a unified derivation of these different types of
covariant time derivatives.  We do not restrict ourselves to Eulerian and
Lagrangian coordinates, as this obscures the general covariance of the theory:
both these descriptions lack certain terms that vanish because of the special
nature of the coordinates.  From a dynamical system defined in some Eulerian
frame, we transform to general time-dependent coordinates.  We then find a
transformation law between two time-dependent frames with no explicit
reference to the Eulerian coordinates.  The Eulerian velocity of the flow is
not a tensor, but the move to general coordinates allows the identification of
a \emph{velocity tensor} that transforms appropriately (\secref{timedep}).  We
also derive a time evolution equation for the Jacobian matrix of a coordinate
transformation between two arbitrary time-independent frames.  This time
evolution equation facilitates the construction of the covariant time
derivative in \secref{covtderiv}.  After a discussion of the \rostr\
tensor in \secref{deftens}, we present in \secref{3derivs} the three types of
covariant time derivatives mentioned above: convective, corotational, and
directional.

\secref{timecurv} addresses a fundamental issue when dealing with generalised
coordinates: the problem of commuting derivatives.  In manipulating fluid
equations it is often necessary to commute the order of time and space
derivation.  When commuting two covariant spatial derivatives, the Riemann
curvature tensor must be taken into account.  Similarly, when commuting a
covariant time derivative with a spatial derivative, there arises a tensor we
call the \emph{time-curvature}.  This tensor vanishes for sufficiently simple
time-dependence of the metric, and satisfies many properties similar to the
Riemann tensor.

Throughout this paper, we will usually refer to the ``fluid,'' ``fluid
elements,'' and ``velocity,'' but this is merely a useful concretion.  The
methods developed apply to general dynamical systems where the velocity is
some arbitrary vector field defined on a manifold.  The covariant time
derivative still refers to the rate of change of tensors along the trajectory,
but the tensors do not necessarily correspond to identifiable physical
quantities.  For example, the covariant time derivative is useful in
fornmulating methods for finding Lyapunov exponents on manifolds with
nontrivial metrics~\cite{Thiffeault2001d}.

\section{Time-dependent Coordinates}
\seclabel{timedep}

We consider the dynamical system on an~$\sdim$-dimensional smooth
manifold~$\Manif$,
\begin{equation}
	\dot\xv = \vel(\time,\xv),
	\eqlabel{dynsys}
\end{equation}
where the overdot indicates a time derivative and~$\vel$ is a differentiable
vector field.  (For simplicity, we restrict ourselves to a given chart.)  A
solution~$\xv(\time)$ defines a curve~$\mathcal{C}$ in~$\Manif$ with
tangent~$\vel$.  We view the~$\xv$ as special coordinates, called the Eulerian
coordinates, and denote vectors expressed in the Eulerian coordinate
basis~$\{\pd/\pd\xv^i\}$ by the indices~$i,j,k$.

A time-dependent coordinate change~$\zv(\time,\xv)$ satisfies
\begin{equation}
	\dot\z^{a}(\time,\xv(\time))
	= \frac{\pd\z^{a}}{\pd\xv^k}\,\vel^k
		+ \frac{\pd\z^{a}}{\pd\time}\holdconst{\x}\,,
	\eqlabel{zdottrans}
\end{equation}
where the~$\pd/\pd\time\sholdconst{\x}$ is taken at constant~$\x$.  Here and
throughout the rest of the paper, we assume the usual Einstein convention of
summing over repeated indices.  We denote vectors expressed in the general
coordinate basis~$\{\pd/\pd\zv^a\}$ by the indices~$a,b,c,d$.  We use the
shorthand notation that the index on a vector~$\vvec$ characterises the
components of that vector in the corresponding basis: thus~$\vvec^a$
and~$\vvec^i$ are the components of~$\vvec$ in the bases~$\{\pd/\pd\zv^a\}$
and~$\{\pd/\pd\x^i\}$, respectively.  The components~$\vvec^a$ and~$\vvec^i$
are also understood to be functions of~$\zv$ and~$\x$, respectively, in
addition to depending explicitly on time.

Defining~$\vel\ldef\dot\zv$, we can regard Eq.~\eqref{zdottrans} as a
transformation law for~$\vel$,
\begin{equation}
	\vel^{a}
	= \frac{\pd\z^{a}}{\pd\xv^k}\,\vel^k
		+ \frac{\pd\z^{a}}{\pd\time}\holdconst{\x}.
	\eqlabel{veltransx}
\end{equation}
This last term prevents~$\vel$ from transforming like a tensor. (We refer the
reader to standard texts in differential geometry for a more detailed
discussion of tensors~\cite{Misner,Schutz,Wald}.)

Now consider a second coordinate system~$\bar\zv(\time,\xv)$, also defined in
terms of~$\xv$.  We can use Eq.~\eqref{veltransx} and the chain rule to define
a transformation law between~$\zv$ and~$\bar\zv$,
\begin{equation}
	\vel^{a} - \frac{\pd\z^{a}}{\pd\time}\holdconst{\x}
	= \frac{\pd\z^{a}}{\pd{\bar\zv}^{\bar a}}\l(
		{\vel}^{\bar a} - \frac{\pd{\bar\z}^{\bar a}}{\pd\time}
		\holdconst{\x}\r).
	\eqlabel{veltrans}
\end{equation}
Any explicit reference to the coordinates~$\xv$ has disappeared (except
in~$\pd/\pd\z\sholdconst{\x}$).  Equation~\eqref{veltrans} is a transformation
law between any two coordinate systems defined in terms of~$\xv$, and implies
that~\hbox{$\vel^{a} - ({\pd\z^{a}}/{\pd\time})\sholdconst{\x}$} transforms
like a tensor.  This suggests defining the tensor
\begin{equation}
	\Vel^a \ldef \vel^a - \frac{\pd\z^a}{\pd\time}\holdconst{\x},
	\eqlabel{Veldef}
\end{equation}
which we call the velocity tensor.  The velocity tensor is the absolute
velocity of the fluid~$\vel$ with the velocity of the coordinates subtracted.

In addition to the coordinates~$\xv$, characterised
by~\hbox{$\pd\xv^i/\pd\time\sholdconst{\x}=0$}, we introduce another special
set of coordinates, the Lagrangian coordinates~$\lagrcv$, defined
by~\hbox{$\dot\lagrcv=0$}.  We denote vectors expressed in the Lagrangian
coordinate basis~$\{\pd/\pd\lagrc^q\}$ by the indices~$p$ and~$q$. From
Eq.~\eqref{veltransx}, we have,
\begin{equation}
	\vel^q(\time,\xv(\time))
		= \frac{\pd\lagrc^q}{\pd\xv^k}\,\vel^k
		+ \frac{\pd\lagrc^q}{\pd\time}\holdconst{\x} = 0.
	\eqlabel{lagrcdef}
\end{equation}
The initial conditions for~$\lagrcv$ are chosen such that Eulerian and
Lagrangian coordinates coincide at~$\time=0$:~$\lagrcv(0,\xv)=\xv$.

Lagrangian and Eulerian coordinates have the advantage that the time evolution
of their Jacobian matrix is easily obtained.  The Jacobian
matrix~${\pd\x^i}/{\pd\lagrc^q}$ satisfies~\cite{Oldroyd1950}
\begin{equation}
	\frac{d}{d\time}\l(\frac{\pd\x^i}{\pd\lagrc^q}\r)
	= \frac{\pd\vel^i}{\pd\x^k}\,\frac{\pd\x^k}{\pd\lagrc^q}.
	\eqlabel{jacevolEulLag}
\end{equation}
By using the identity
\begin{equation*}
	\frac{d}{d\time}\l(\frac{\pd\x^i}{\pd\lagrc^q}\,
		\frac{\pd\lagrc^p}{\pd\x^i}\r)
	 = \frac{d}{d\time}\l({\delta_q}^p\r) = 0,
\end{equation*}
which follows from the chain rule, and using the Leibniz property and
Eq.~\eqref{jacevolEulLag}, we find
\begin{equation*}
	\frac{d}{d\time}\l(\frac{\pd\lagrc^q}{\pd\x^i}\r)
	= -\frac{\pd\lagrc^q}{\pd\x^k}\,\frac{\pd\vel^k}{\pd\x^i}.
\end{equation*}
The Leibniz property can be used again to find the time evolution of the
Jacobian matrix of two arbitrary time-dependent
transformations~$\z(\time,\xv)$ and~$\bar\z(\time,\xv)$,
\begin{equation}
	\frac{d}{d\time}\l(\frac{\pd\z^a}{\pd\bar\z^{\bar a}}\r)
	= \frac{\pd\vel^a}{\pd\z^b}\,\frac{\pd\z^b}{\pd\bar\z^{\bar a}}
	- \frac{\pd\z^a}{\pd\bar\z^{\bar b}}\,
		\frac{\pd\vel^{\bar b}}{\pd\bar\z^{\bar a}}.
	\eqlabel{jacevol}
\end{equation}
All reference to Eulerian and Lagrangian coordinates has disappeared from
Eq.~\eqref{jacevol}; this equation is crucial when deriving the covariant
time derivative of \secref{covtderiv}.

\section{The Covariant Time Derivative}
\seclabel{covtderiv}

The standard time derivative operator, which we have been denoting by an
overdot, is defined for a vector field~$\vvec$ as
\begin{equation}
	\dot\vvec^a \ldef \frac{\pd\vvec^a}{\pd\time}\holdconst{\z}
		+ \frac{\pd\vvec^a}{\pd\z^b}\,\vel^b,
	\eqlabel{dotdef}
\end{equation}
where recall that~$\dot\zv=\vel$.  The first term is the change in~$\vvec$ due
to any explicit time-dependence it might have; the second term is the change
in~$\vvec$ due to its dependence on~$\zv$.  The time derivative is not
covariant, because a time-dependent change of basis will modify the form of
Eq.~\eqref{dotdef}.

We define the covariant time derivative~$\Dt{}$ by
\begin{equation}
	\Dt{\vvec^a} \ldef \dot\vvec^a + {\ptconnex^a}_b\,\vvec^b\,,
	\eqlabel{Dtdef}
\end{equation}
where the~${\ptconnex^a}_b$ are time-dependent quantities that are chosen to
make~$\Dt\vvec^a$ covariant.  In order that the operator~$\Dt$ have the
Leibniz property, and that it reduce to the ordinary derivative~\eqref{dotdef}
when acting on scalars, we require
\begin{equation*}
	\Dt{\wvec_a} = \dot\wvec_a - {\ptconnex^b}_a\,\wvec_b\,,
\end{equation*}
when acting on a~1-form~$\wvec$.  When~$\Dt$ acts on mixed tensors of higher
rank, an~$\ptconnex$ must be added for each superscript, and one must be
subtracted for each subscript.  We refer to the~$\ptconnex$ as
\emph{connexions}, by analogy with the spatial derivative case.

By enforcing covariance of~$\Dt$, we can derive a general expression
for~${\ptconnex^a}_b$.  Since~$\vvec$ is a tensor, we can write
\begin{align*}
	\Dt{\vvec^a} &=
	\Dt{\l(\frac{\pd\z^a}{\pd\bar\z^{\bar a}}\,\vvec^{\bar a}\r)}\\
	&= \frac{d}{d\time}
		{\l(\frac{\pd\z^a}{\pd\bar\z^{\bar a}}\r)}\vvec^{\bar a}
	+ \frac{\pd\z^a}{\pd\bar\z^{\bar a}}\,\dot{\vvec}^{\bar a}
	+ \frac{\pd\z^b}{\pd\bar\z^{\bar a}}\,
		\vvec^{\bar a}\,{\ptconnex^a}_b\\
	&= \frac{\pd\z^a}{\pd\bar\z^{\bar a}}\,\Dt{\vvec^{\bar a}},
\end{align*}
because~$\Dt{\vvec^a}$ is by definition covariant.
Hence, we require
the~$\ptconnex$'s to transform as
\begin{equation}
	{\ptconnex^{\bar a}}_{\bar b}
	= \frac{\pd\bar\z^{\bar a}}{\pd\z^{a}}\,
		\frac{\pd\z^{b}}{\pd\bar\z^{\bar b}}\,{\ptconnex^{a}}_{b}
	+ \frac{\pd\bar\z^{\bar a}}{\pd\z^c}\,
		\frac{d}{d\time}\l(\frac{\pd\z^c}{\pd\bar\z^{\bar b}}\r).
	\eqlabel{ptconnextrans}
\end{equation}
The first term in~\eqref{ptconnextrans} is the usual tensorial transformation
law.  The second term implies that~$\ptconnex$ is not a tensor, and arises
because of the time-dependence.

Inserting the evolution Eq.~\eqref{jacevol} into~\eqref{ptconnextrans}, we can
rewrite the transformation law for~$\ptconnex$ as
\begin{equation}
	{\ptconnex^{\bar a}}_{\bar b}
		+ \frac{\pd\vel^{\bar a}}{\pd{\bar\z}^{\bar b}}
	= \frac{\pd\bar\z^{\bar a}}{\pd\z^{a}}\,
		\frac{\pd\z^{b}}{\pd\bar\z^{\bar b}}\l({\ptconnex^{a}}_{b}
		+ \frac{\pd\vel^{a}}{\pd\z^{b}}\r),
	\eqlabel{ptconnextrans2}
\end{equation}
implying that~${\ptconnex^a}_b+(\pd\vel^a/\pd\z^b)$ transforms like a tensor.
Hence,
\begin{equation}
	{\ptconnex^a}_b = -\frac{\pd\vel^a}{\pd\z^b}
		+ {\anytens^a}_b\,,
	\eqlabel{ptconnexgeneral}
\end{equation}
where~$\anytens$ is an arbitrary tensor.  Equation~\eqref{ptconnexgeneral} is
the most general form of the connexions~$\ptconnex$.

In~\secref{3derivs}, we consider three convenient choices of the
tensor~$\anytens$.  But first in~\secref{deftens} we examine the action of the
covariant derivative on the metric tensor.

\section{The Rate-of-strain Tensor}
\seclabel{deftens}

Our development so far has not made use of a metric tensor.  We now introduce
such a tensor, specifically a Riemannian
metric~$\metric:\Tangent\Manif\times\Tangent\Manif\rightarrow\Re$.  The
components~$\metric_{ab}$ of the metric are functions of~$\z$ and~$\time$, but
the indices~$a$ and~$b$ run over the dimension~$\sdim$ of~$\Tangent\Manif$,
and so do not include a time component.

It is informative to consider the derivative of the metric tensor,
\begin{align*}
	\Dt\metric_{ab} &= \dot\metric_{ab} - {\ptconnex^c}_a\,\metric_{bc}
		- {\ptconnex^c}_b\,\metric_{ac}\\
	&= \frac{\pd\metric_{ab}}{\pd\time}\holdconst{\z}
		+ \frac{\pd\metric_{ab}}{\pd\z^c}\,\vel^c
		+ \metric_{ac}\,\frac{\pd\vel^c}{\pd\z^b}
		+ \metric_{bc}\,\frac{\pd\vel^c}{\pd\z^a}
		- (\anytens_{ab} + \anytens_{ba}),
\end{align*}
where we have used the metric to lower the indices on~$\anytens$.  We define
the intrinsic \emph{\rostr} or \emph{rate-of-deformation}
tensor~$\rostrain$~\cite{Scriven1960,Aris} as
\begin{equation}
	\rostrain_{ab}
	= \half\l[
	\metric_{ac}\covd_b\vel^c + \metric_{bc}\covd_a\vel^c
	+ \frac{\pd\metric_{ab}}{\pd\time}\holdconst{\z}\r].
	\eqlabel{rostraindef}
\end{equation}
Here we denote by~$\covd_a$ the covariant derivative with respect to~$\z^a$,
\begin{equation*}
	\covd_b\vvec^a \ldef \frac{\pd\vvec^a}{\pd\z^b}
		+ \Connex^a_{bc}\,\vvec^c\,.
\end{equation*}
The Riemann--Christoffel connexions are defined as~\cite{Schutz,Wald}
\begin{equation}
	\Connex^a_{bc} \ldef \half\metric^{ad}\l(
	\frac{\pd\metric_{bd}}{\pd\z^c}
	+ \frac{\pd\metric_{cd}}{\pd\z^b}
	- \frac{\pd\metric_{bc}}{\pd\z^d}\r),
	\eqlabel{Connexdef}
\end{equation}
whence the identity
\begin{equation*}
      \metric_{ac}\Connex^c_{bd}
      + \metric_{bc}\Connex^c_{ad} = \frac{\pd\metric_{ab}}{\pd\z^d}
\end{equation*}
holds.  The covariant time derivative of the metric can thus be rewritten
\begin{equation}
	\half\Dt\metric_{ab}
	= \rostrain_{ab} - \anytenssym_{ab}\,,
	\eqlabel{Dtmetric}
\end{equation}
where~$\anytenssym\ldef\half(\anytens_{ab}+\anytens_{ba})$ denotes the
symmetric part of~$\anytens$.

The \rostr\ tensor~$\rostrain$ describes the stretching of fluid elements.
The time derivative of the metric in its definition~\eqref{rostraindef} is
necessary for covariance under time-dependent transformations; the term
describes straining motion that is inherent to the space, as embodied by the
metric.  The trace of the \rostr\ tensor is a scalar
\begin{equation}
	{\rostrain^c}_c = \metric^{ac}\,\rostrain_{ac} = \covd_c\,\vel^c
		+ \half\frac{\pd}{\pd\time}\holdconst{\z}\log|\metric|\,,
	\eqlabel{rostraintrace}
\end{equation}
where~$|\metric|$ is the determinant of~$\metric_{ab}$ and we have used the
identity
\begin{equation}
	\metric^{ac}\,\frac{\pd\metric_{ac}}{\pd\time}\holdconst{\z}
		= \frac{\pd}{\pd\time}\holdconst{\z}\log|\metric|\,.
\end{equation}
The \rostr\ tensor can be decomposed as
\begin{equation}
	\rostrain_{ab} = \rostrain'_{ab} +
		\frac{1}{\sdim}\,{\rostrain^c}_c\,\metric_{ab}\,,
\end{equation}
where~$\rostrain'_{ab}$ is traceless and represents a straining motion without
change of volume, and~${\rostrain^c}_c\,\metric_{ab}/\sdim$ is an isotropic
expansion.  We see from the trace~\eqref{rostraintrace} that for a
time-dependent metric there can be an isotropic expansion even for an
incompressible flow, if~$\pd|\metric|/\pd\time\sholdconst{\z}\ne0$.  Note also
that in Lagrangian coordinates (characterised by $\vel^q=0$), the \rostr\
tensor reduces to
\begin{equation*}
	\rostrain_{pq}
	= \half\frac{\pd\metric_{pq}}{\pd\time}\holdconst{\lagrcv}\,,
\end{equation*}
so that the deformation of the space is contained entirely in the metric
tensor.

\section{Three Covariant Derivatives}
\seclabel{3derivs}

As mentioned in \secref{timedep}, the requirement of covariance only fixes the
covariant time derivative up to an arbitrary tensor
[Eq.~\eqref{ptconnexgeneral}].  That tensor may be chosen to suit the problem
at hand, but there are three particular choices that merit special attention.
In \secref{dconvective} we treat the convective derivative, and in
\secref{compatd} we examine two types of compatible derivatives: corotational
and directional.

\subsection{The Convective Derivative}
\seclabel{dconvective}

The choice~\hbox{$\anytens\equiv 0$} is equivalent to the convective
derivative of Oldroyd~\cite{Oldroyd1950,Aris}.  The connexion,
Eq.~\eqref{ptconnexgeneral}, reduces to the simple form
\begin{equation*}
	{\ptconnex^a}_b = - \frac{\pd\vel^a}{\pd\z^b}.
\end{equation*}
The convective derivative~$\Dtconv$ acting on a vector~$\vvec^a$ is thus
\begin{equation}
	\Dtconv\vvec^a = \frac{\pd\vvec^a}{\pd\time}\holdconst{\z}
		+ \frac{\pd\vvec^a}{\pd\z^b}\,\vel^b
		- \frac{\pd\vel^a}{\pd\z^b}\,\vvec^b\,.
	\eqlabel{Dtconvvec}
\end{equation}
When acting on a contravariant vector~$\vvec^a$, as in Eq.~\eqref{Dtconvvec},
$\Dtconv$ is sometimes called the \emph{upper convected}
derivative~\cite{Jou}; $\Dtconv$ acting on a covariant
vector~$\wvec_a$, \hbox{$\Dtconv\wvec_a = \dot\wvec_a +
(\pd\vel^b/\pd\z^a)\,\wvec_b$}, is then called the \emph{lower convected}
derivative.

In general, for an arbitrary tensor~$\arbtens$,
\begin{equation*}
	\Dtconv\arbtens = \frac{\pd\arbtens}{\pd\time}\holdconst{\z}
		+ \mathcal{L}_{\vel}\arbtens,
\end{equation*}
where~$\mathcal{L}_{\vel}\arbtens$ is the Lie derivative of~$\arbtens$ with
respect to~$\vel$~\cite{Schutz,Wald}.  In Lagrangian coordinates, we
have~$\vel^q\equiv0$, so the convective derivative reduces to
\begin{equation*}
	\Dtconv\vvec^q = \frac{\pd\vvec^q}{\pd\time}\holdconst{\lagrcv}.
\end{equation*}

The convective derivative is not compatible with the metric: from
Eq.~\eqref{Dtmetric}, the metric's derivative is
\begin{equation*}
	\half\Dtconv\metric_{ab} = \rostrain_{ab}\,.
\end{equation*}
which does not vanish, unless the velocity field is strain-free.

The convective derivative is ideally suited to problems of \emph{advection}
with stretching, where a tensor is carried and stretched by a velocity field.
Table~\ref{tab:advec} summarises the form of the equation for advection with
stretching of a vector field~$\Bf$ ($\Bf$ is ``frozen in'' the
flow~\cite{STF}) for the three different types of derivatives introduced here.
The equation for the contravariant component~$\Bf^a$ is
simply~\hbox{$\Dtconv\Bf^a=0$}, but the equation for the covariant
component~\hbox{$\Bf_a=\metric_{ac}\,\Bf^c$}
is~\hbox{$\Dtconv\Bf_a=2{\Bf_c}\,{\rostrain^c}_a$}.  These two equations
differ because the operator~$\Dtconv$ is not compatible with the metric.

\begin{table}
\caption{Comparison of the equation of motion for the components of an
advected and stretched vector field~$\Bf$.  The equations for the covariant
and contravariant components of~$\Dtconv\Bf$ differ because of the lack of
compatibility with the metric.}
\label{tab:advec}
\vspace{1em}
\begin{center}
\begin{tabular}{lll}
\textsl{Type} &
\textsl{Contravariant components} & \textsl{Covariant components}\\
\hline
Convective &
$\Dtconv\Bf^a=0$ & $\Dtconv\Bf_a=2\Bf_c\,{\rostrain^c}_a$\\
Corotational &
$\Dtcor\Bf^a=\Bf^c\,{\rostrain_c}^a$ &
$\Dtcor\Bf_a=\Bf_c\,{\rostrain_a}^c$\\
Directional &
$\Dtdir\Bf^a=\Bf^c\,(\covd_c\,\Vel^a + {\rocstrain_c}^a)$ &
$\Dtdir\Bf_a=\Bf_c\,(\covd^c\,\Vel_a + {\rocstrain_a}^c)$ \\
\hline
\end{tabular}
\end{center}
\vspace{2em}
\end{table}

\subsection{Compatible Derivatives}
\seclabel{compatd}

Another way to fix~$\anytens$ is to require that the operator~$\Dt$ be
compatible with the metric, that is,~$\Dt\metric_{ab}=0$.  This allows us to
raise and lower indices through the operator~$\Dt$, a property possessed by
the covariant spatial derivative.  From Eq.~\eqref{Dtmetric}, the
requirement~$\Dt\metric_{ab}=0$ uniquely specifies the symmetric part
of~$\anytens_{ab}$, so that~$\anytenssym=\rostrain$.  Using
Eqs.~\eqref{ptconnexgeneral} and~\eqref{rostraindef}, we then find
\begin{equation}
	\ptconnex_{ab} = \metric_{ac}\,\Connex^c_{bd}\,\vel^d
	+ \half\frac{\pd\metric_{ab}}{\pd\time}\holdconst{\z}
	- \half\l[\metric_{ac}\covd_b\vel^c - \metric_{bc}\covd_a\vel^c\r]
	+ \anytensasym_{ab}\,,
	\eqlabel{ptconnexcompat}
\end{equation}
where~$\anytensasym\ldef\half(\anytens_{ab}-\anytens_{ba})$ is the
antisymmetric part of~$\anytens$.

We define the antisymmetric vorticity tensor
\begin{equation}
	\Vortens_{ab} \ldef
	\half\l[\metric_{ac}\,\frac{\pd\Vel^c}{\pd\z^b}
	- \metric_{bc}\,\frac{\pd\Vel^c}{\pd\z^a}\r]
	\eqlabel{Vortdef}
\end{equation}
and the symmetric \crostr\ tensor
\begin{equation}
	\rocstrain_{ab} \ldef
	\half\l[\metric_{ac}\covd_b\l(\frac{\pd\z^c}{\pd\time}\holdconst{\x}\r)
	+ \metric_{bc}\covd_a\l(\frac{\pd\z^c}{\pd\time}\holdconst{\x}\r)
		+ \frac{\pd\metric_{ab}}{\pd\time}\holdconst{\z}\r].
	\eqlabel{rocstraindef}
\end{equation}
In Eulerian coordinates, we
have~\hbox{$\rocstrain_{ij}=\half(\pd\metric_{ij}/\pd\time)\sholdconst{\x}$}. The
compatible connexion~\eqref{ptconnexcompat} can be rewritten
\begin{equation}
	\ptconnex_{ab} = \metric_{ac}\,\Connex^c_{bd}\,\vel^d
	- \metric_{ac}\covd_b\l(\frac{\pd\z^c}{\pd\time}\holdconst{\x}\r)
	+ \rocstrain_{ab}
	- \Vortens_{ab}
	+ \anytensasym_{ab}\,.
	\eqlabel{ptconnexcompat2}
\end{equation}
Since~$\anytensasym_{ab}$ is antisymmetric, we can use it to cancel the
vorticity, or we can set it to zero.  The two choices are discussed separately
in \secreftwo{corot}{dcurve}.

The decomposition of the velocity gradient tensor~$\covd\Vel$ into the \rostr\
and vorticity tensors has the form
\begin{equation}
	\metric_{ac}\,\covd_b\Vel^c
	= \l[\rostrain_{ab} - \rocstrain_{ab}\r] + \Vortens_{ab}
\end{equation}
in general time-dependent coordinates.  When the coordinates have no time
dependence, the tensor~$\rocstrain$ vanishes, as does the
derivatives~$\pd\z^c/\pd\time\sholdconst{\x}$, and we recover the usual
decomposition of the velocity gradient tensor into the \rostr\ and the
vorticity.  We can think of~$\rocstrain$ as the contribution to the \rostr\
tensor that is due to coordinate deformation and not to gradients of the
velocity field.  However, the term~$\pd\metric/\pd\time\sholdconst{\z}$ is a
``real'' effect representing the deformation due to a time-dependent metric,
and is thus also included in the definition of the intrinsic \rostr\
tensor,~$\rostrain$, defined by Eq.~\eqref{rostraindef}.

In Euclidean space, when the \rostr\ tensor~$\rostrain$ vanishes everywhere we
are left with rigid-body rotation at a constant rate given
by~$\Vortens$~\cite{Aris}.  With an arbitrary metric and time-dependent
coordinates the situation is not so simple: the very concept of rigid-body
rotation is not well-defined.  Hence, even when~\hbox{$\rostrain\equiv 0$}, we
cannot expect to be able to solve for~$\vel$ in closed form.

\subsubsection{The Corotational Derivative}
\seclabel{corot}

In this instance we choose the antisymmetric part~$\anytensasym_{ab}$ to be
zero.  We call \emph{corotational} the resulting covariant derivative, and
denote it by~$\Dtcor$ (the subscript~$J$ stands for \emph{Jaumann}).  The
appellation ``corotational'' really applies to the Euclidean
limit,~\hbox{$\metric_{ij}=\delta_{ij}$}, for which the compatible connexion
Eq.~\eqref{ptconnexcompat} reduces to~\hbox{$\ptconnex_{ij}=-\Vortens_{ij}$}.
It is then clear that the covariant derivative is designed to include the
effects of local rotation of the flow, as embodied by the vorticity.  (See
Refs.~\cite{Jou} and~\cite[p.~342]{Bird1}, and references therein.)  The
derivative~\eqref{ptconnexcompat} with~$\anytensasym\equiv0$ is thus a
generalisation of the corotational derivative to include the effect of
time-dependent non-Euclidean coordinates.

In Table~\ref{tab:advec}, we can see that, written using~$\Dtcor$, the
equation for advection with stretching of a vector~$\Bf^a$ has the \rostr\
tensor on the right-hand side.  The ``rotational'' effects are included
in~$\Dtcor$, hence the terms that remain include only the strain.

\subsubsection{The Directional Derivative}
\seclabel{dcurve}

Another convenient choice is to set~$\anytensasym_{ab}=\Vortens_{ab}$, thus
cancelling the vorticity in Eq.~\eqref{ptconnexcompat2}.  The resulting
covariant time derivative then has the property that, in the absence of any
explicit time-dependence, it reduces to the covariant derivative along the
curve~$\mathcal{C}$~\cite{Schutz,Wald}, or directional derivative,
where~$\mathcal{C}$ is the trajectory of the dynamical system in the general
coordinates~$\z$ (\secref{timedep}).  The derivative is called directional
because it only depends on~$\vel$, and not gradients of~$\vel$.

The form of the equation for advection with stretching of a vector~$\Bf^a$
written using~$\Dtdir$ is shown in Table~\ref{tab:advec}.  The~$\covd\Vel$
term on the right-hand side is the ``stretching'' term~\cite{STF} (called
\emph{vortex stretching} when~$\Bf$ is the vorticity vector~\cite{Tritton}).
The~$\rocstrain$ term represents coordinate stretching, and does not appear in
Euclidean space with time-independent coordinates.

Because the directional derivative depends only on~$\vel$ and not its
gradients, it can be used to define time-dependent parallel transport of
tensors.  A vector~$\vvec$ is said to be \emph{parallel transported}
along~$\vel$ if it satisfies~\hbox{$\Dtdir\,\vvec = 0$}, or equivalently
\begin{equation}
	\frac{\pd\vvec^a}{\pd\time}\holdconst{\z} + \vel^c\,\covd_c\vvec^a
		= \vvec^c\l[\covd_c\l(
			\frac{\pd\zv^a}{\pd\time}\holdconst{\x}\r)
		- {\rocstrain^a}_c\r].
	\eqlabel{partransexpl}
\end{equation}
This can be readily generalised to tensors of higher rank.  In Euclidean
space, with time-independent coordinates, the right-hand side of
Eq.~\eqref{partransexpl} vanishes, leaving only advection of the components
of~$\vvec$.  Thus, parallel transport is closely related to advection without
stretching; Equation~\eqref{partransexpl} is the covariant formulation of the
passive advection equation.

\section{Time-curvature}
\seclabel{timecurv}

A hallmark of generalised coordinates is the possibility of having nonzero
curvature.  The curvature reflects the lack of commutativity of covariant
derivatives, and is tied to parallel transport of vectors along
curves~\cite{Schutz,Wald}.  An analogous curvature arises when we try to
commute~$\Dt$ and~$\covd$, respectively the covariant time and space
derivatives:
\begin{multline}
	\covd_a[\Dt{\vvec^b}] - \Dt{[\covd_a\vvec^b]}
	= {\anytens^c}_a\,\covd_c\vvec^b\\
	+ \metric^{bc}\Bigl[
		\covd_a\l(\anytens_{cd} - \rostrain_{cd} - \Vortens_{cd}\r)
		+ \Rcurv_{cdae}\Vel^e
		+ \half\,\Scurv_{cda}\Bigr]\vvec^d\,,
	\eqlabel{commute}
\end{multline}
where the time-curvature tensor is defined by
\begin{multline}
	\Scurv_{abc} \ldef
	\covd_a\l[\frac{\pd\metric_{cb}}{\pd\time}\holdconst{\z}
		+ \metric_{be}\covd_c\l(
			\frac{\pd\z^e}{\pd\time}\holdconst{\x}\r)\r]\\
	- \covd_b\l[\frac{\pd\metric_{ca}}{\pd\time}\holdconst{\z}
		+ \metric_{ae}\covd_c\l(
			\frac{\pd\z^e}{\pd\time}\holdconst{\x}\r)\r]
	+ \Rcurv_{abce}\,\frac{\pd\z^e}{\pd\time}\holdconst{\x}\,,
	\eqlabel{Scurvdef}
\end{multline}
and the Riemann curvature tensor~$\Rcurv$ obeys~\cite{Wald}
\begin{equation}
	(\covd_c\covd_d - \covd_d\covd_c)\,\vvec^a
	= {\Rcurv^a}_{bcd}\,\vvec^b.
	\eqlabel{Rcurvdef}
\end{equation}
The time-curvature tensor satisfies~\hbox{$\Scurv_{abc}=-\Scurv_{bac}$}, and
the Riemann curvature tensor satisfies~\hbox{$\Rcurv_{abcd} =
-\Rcurv_{bacd}$},~\hbox{$\Rcurv_{abcd} = \Rcurv_{cdab}$}.

Even for trivial (Euclidean) coordinates, we do not expect~$\Dt$ and~$\covd$
to commute, because of the derivatives of~\hbox{$\vel$ in
the~$\covd_a\l(\anytens_{cd} - \rostrain_{cd} - \Vortens_{cd}\r)$} term of
Eq. \eqref{commute}.  Note that the \crostr\ tensor~$\rocstrain$, defined by
Eq.~\eqref{rocstraindef}, does not appear in Eq.~\eqref{commute}.

The~$\covd\vvec$ term in Eq.~\eqref{commute} vanishes for the convective
derivative of \secref{dconvective}, since then~$\anytens\equiv0$.  For the
directional derivative of \secref{dcurve}, we have \hbox{$\anytens_{cd} =
\rostrain_{cd} + \Vortens_{cd}$}, so the commutation relation simplifies to
\begin{equation*}
	\covd_a[\Dt{\vvec^b}] - \Dt{[\covd_a\vvec^b]}
	= \l(\rostrain_{cd} + \Vortens_{cd}\r)\covd_c\vvec^b
	+ \metric^{bc}\Bigl[
		\Rcurv_{cdae}\Vel^e
		+ \half\,\Scurv_{cda}\Bigr]\vvec^d,
\end{equation*}
which does not involve second derivatives of~$\vel$.  For the corotational
derivative of \secref{corot}, with~$\anytens_{cd}=\rostrain_{cd}$, no terms
drop out.

The terms involving~$\anytens$ in the commutation relation~\eqref{commute}
reflect properties of the velocity field~$\vel$.  In contrast, the
tensors~$\Rcurv$ and~$\Scurv$ embody intrinsic properties of the metric
tensor~$\metric$.  The Riemann tensor~$\Rcurv$ is nonzero when the space is
curved.  The time-curvature tensor~$\Scurv$ is new and has characteristics
that are analogous to the Riemann tensor.  It satisfies a cyclic permutation
identity,
\begin{equation}
	\Scurv_{abc} + \Scurv_{cab} + \Scurv_{bca} = 0,
	\eqlabel{Bianchi1}
\end{equation}
which corresponds to the first Bianchi identity of the Riemann tensor.  The
time-curvature does not appear to satisfy an analogue of the second Bianchi
identity.

The property~\hbox{$\Scurv_{abc}=-\Scurv_{bac}$}, together with the Bianchi
identity~\eqref{Bianchi1}, imply that $\Scurv$ has \hbox{$\sdim(\sdim^2-1)/3$}
independent components, compared to the \hbox{$\sdim^2(\sdim^2-1)/12$}
components of~$\Rcurv$, where~$\sdim$ is the dimension of the space.  Thus
one-dimensional manifolds have vanishing~$\Scurv$ and~$\Rcurv$.
For~$1\le\sdim\le3$,~$\Scurv$ has more independent components than~$\Rcurv$;
for~$\sdim=4$, they both have~$20$.  For~$\sdim>4$,~$\Rcurv$ has more
independent components than~$\Scurv$.

The time-curvature~$\Scurv$ vanishes for a time-independent metric and
coordinates.  It also vanishes for a metric of the
form~$\metric_{ij}(\time,\xv) = \beta(\time)\,h_{ij}(\xv)$, where~$h$ is a
time-independent metric and~$\xv$ are the Eulerian coordinates.  It follows
from its tensorial nature that the time-curvature must then vanish in any
time-dependent coordinates.  In general, it is convenient to find~$\Scurv$ in
Eulerian coordinates (where $\pd\x/\pd\time\sholdconst{\x}=0$),
\begin{equation}
	\Scurv_{ijk} \ldef
	\covd_i\l(\frac{\pd\metric_{kj}}{\pd\time}\holdconst{\z}\r)
	- \covd_j\l(\frac{\pd\metric_{ki}}{\pd\time}\holdconst{\z}\r)\,,
	\eqlabel{ScurvEul}
\end{equation}
and then transform~$\Scurv_{ijk}$ to arbitrary time-dependent coordinates
using the tensorial law.

\section{Discussion}

In this paper, we aimed to provide a systematic framework to handle
complicated time-dependent metrics and coordinate systems on manifolds.  The
explicit form of the relevant tensors is often fairly involved, but the
advantage is that they can be evaluated in time-independent Eulerian
coordinates and then transformed to arbitrary coordinate systems using the
usual tensorial transformation laws.

The covariance of the time derivatives is made explicit by using arbitrary
time-dependent coordinates.  The results for the Eulerian coordinates~$\x^i$
are recovered by setting~$\pd\x^i/\pd\time\sholdconst{\x}=0$, and those for
the Lagrangian coordinates~$\lagrcv^q$ by setting~$\vel^q=0$.

The introduction of the time-curvature tensor allows us to treat the temporal
dependence of the metric tensor in a manner analogous to its spatial
dependence.  For simple time-dependence, the time-curvature vanishes, such as
for the case of a time-independent metric multiplied by a time-dependent
scalar.  As for the (spatial) Riemann curvature tensor, the components of the
time-curvature can be computed for a given metric, and then inserted whenever
a temporal and spatial derivative need to be commuted.

We have only addressed the \emph{kinematics} of fluid motion.  The dynamical
equations relating the rate of change of quantities to the forces in play have
not been discussed (see Refs.~\cite{Jou,Oldroyd1950,Scriven1960,Aris}), and
depend on the specifics of the problem at hand.  Nevertheless, covariant time
derivatives provide a powerful framework in which to formulate such dynamical
equations.

\begin{ack}

The author thanks Chris Wiggins for pointing out a useful reference, and Tom
Yudichak for an illuminating discussion.  This work was supported by an NSF/DOE
Partnership in Basic Plasma Science grant, No.~DE-FG02-97ER54441.

\end{ack}


\end{document}